\documentclass[superscriptaddress,secnumarabic,amssymb,amsmath,nobibnotes,aps,prd,showkeys,nofootinbib,preprint]{revtex4}

\usepackage{graphicx}
\usepackage{float}
\usepackage{bm}
\usepackage{amsmath}
\usepackage{amsfonts}
\usepackage{amssymb}
\usepackage{epstopdf}
\usepackage{natbib}%
\newcommand{\bee}{\begin{equation}}
	\newcommand{\eee}{\end{equation}}
\newcommand{\eaa}{\end{eqnarray}}
\newcommand{\baa}{\begin{eqnarray}}

\usepackage{color}

\begin{document}
	
	\title{Modified gauge unfixing formalism and gauge symmetries in the non-commutative chiral bosons theory}
	\author{Cleber N. Costa}
	\email{cleber.costa@ice.ufjf.br}
	\affiliation{Departamento de F\'isica, Universidade Federal de Juiz de Fora, Juiz de Fora - 36036-330, MG, Brazil}
	\author{Gabriella V. Ambrósio}
	\email{gabriellambrosio@gmail.com}
	\affiliation{Departamento de F\'isica, Universidade Federal de Juiz de Fora, Juiz de Fora - 36036-330, MG, Brazil}
	\author{Paulo R. F. Alves}
	\email{paulo.alves@ice.ufjf.br}
	\affiliation{Departamento de F\'isica, Universidade Federal de Juiz de Fora, Juiz de Fora - 36036-330, MG, Brazil}
	\author{Jorge Ananias Neto}
	\email{jorge@fisica.ufjf.br}
	\affiliation{Departamento de F\'isica, Universidade Federal de Juiz de Fora, Juiz de Fora - 36036-330, MG, Brazil}
	\author{Ronaldo Thibes}
	\email{thibes@uesb.edu.br}
	\affiliation{Universidade Estadual do Sudoeste da Bahia, DCEN,
		Rodovia BR 415, Km 03, Itapetinga – 45700-000, Brazil}
	\affiliation{Theoretical Physics Department, CERN, 1211 Geneva 23, Switzerland}

	\begin{abstract}
		We use the gauge unfixing (GU) formalism framework in a two dimensional noncommutative chiral bosons (NCCB) model to disclose new hidden symmetries. That amounts to converting a second-class system to a first-class one without adding any extra degrees of freedom in phase space. The NCCB model has two second-class constraints -- one of them turns out as a gauge symmetry generator while the other one, considered as a gauge-fixing condition, is disregarded in the converted gauge-invariant system. We show that it is possible to apply a conversion technique based on the GU formalism direct to the second-class variables present in the NCCB model, constructing deformed gauge-invariant GU variables, a procedure which we name here as modified GU formalism.  For the canonical analysis in noncommutative phase space, we compute the deformed Dirac brackets between all original phase space variables. 
		We obtain two different gauge invariant versions for the NCCB system and, in each case, a GU Hamiltonian is derived satisfying a corresponding first-class algebra. Finally, the phase space partition function is presented for each case allowing for a consistent functional quantization for the obtained gauge-invariant NCCB.
	\end{abstract}

	\maketitle

	\section{Introduction}\label{sec:Introduction}
	In the last decades,
	chiral bosons (CB) field theory has drawn a lot of interest in the physics community
	\cite{Chizhov:2008tp, Mezincescu:2022hnb, Ouyang:2020rpq, Siegel:1983es, Green:1987sp, bar, Shukla:2013baa, bw, jsonne, Giaccari:2008zx, Bastianelli:1989cu, Wang:2018dua, Wen:1990se, Upadhyay:2011ph, Srivastava:1999ym, Srivastava:2002mw, fj, Amorim:1995tk, McCabe:1990ge, Girotti:1988ua, Rahaman:2016ofe}.   Additionaly to its own relevance, CB have been useful for understanding strings, superstrings and supersymmetry \cite{Siegel:1983es, Green:1987sp, bar, Shukla:2013baa}, gravity and supergravity theories \cite{bw, jsonne, Giaccari:2008zx, Bastianelli:1989cu}, black holes \cite{Wang:2018dua}, fractionary quantum Hall effect \cite{Wen:1990se}, Hodge theory \cite{Upadhyay:2011ph}
	and general aspects of field theories in the light cone \cite{Srivastava:1999ym, Srivastava:2002mw}.
	After the pioneering work by Siegel \cite{Siegel:1983es}, in which the self-dual condition first appeared as a quadratic constraint, Floreaninin and Jackiw introduced a local Lagrangian density for chiral bosons in which a rich canonical structure was revealed \cite{fj}. Despite its initial straightforward simplicity, the description of CB presents some subtleties \cite{Srivastava:2002mw, fj, Amorim:1995tk, McCabe:1990ge, Girotti:1988ua,  Rahaman:2016ofe}. 
	Among several issues, we may mention the presence of a single second-class primary constraint with a non-trivial commutation relation with itself. Therefore CB field theory is not naturally gauge invariant in its original inception.
	As it is well-known, gauge invariant theories hold significant importance regarding its quantum aspects \cite{Thibes:2020jfp}, in particular the quantization of first-class systems is much more easy when compared to second-class theories \cite{mes, Henneaux:1992ig}.
	In the case of CB field theory, obtaining gauge invariance in a consistent form is not a simple task and should be handled with extra care. 
	Due to the presence of an odd number of second-class constraints, in the Floreanini-Jackiw (FJ) description, usual conversional methods such as standard Batalin-Fradkin-Tyutin (BFT) or direct gauge-unfixing do not work.  An alternative successful route can be found in reference \cite{jplb} where, in order to restore CB gauge invariance, a particular method employing both BFT and gauge unfixing ideas was used.
	It is also worth mentioning the constraint Fourier modes expansion approach used in \cite{Amorim:1994np}
	to allow for the introduction of BFT modes.
	
	Over and above that, in the beginning of the current century, there have been various arguments regarding the possibility of noncommutative effects in high energy physics 
	\cite{Douglas:2001ba, Konechny:2000dp, Carmona:2003kh, Carmona:2002iv}.  Relating those two subjects, two extensions of CB field theory including noncommutative features have been proposed in the literature \cite{Miao:2003ab, dgms}.  The first one allows for noncommutativity in space-time coordinates \cite{Miao:2003ab} while the second one introduces noncomutativity directly into the fields space itself \cite{dgms} and has been further investigated in reference \cite{Abreu:2004xe}.  In the present work, we shall be concerned with the latter idea, named here for short as Noncommutative Chiral Bosons (NCCB), in which the quantum operators algebra is deformed in terms of a controlling noncommutative parameter.  The NCCB model, introduced in \cite{dgms}, connects two FJ chiral bosons through that noncommutativity parameter, giving rise to nontrivial commutation relations among the fields in phase space and, similarly to \cite{fj}, is characterized as a constrained second-class system lacking aparent gauge freedom.   Due to the interaction between left and right propagation modes, the original number of constraints doubles, turning more feasible to look for gauge symmetries generated by constraint abelianization methods.  In fact, a couple of recent articles \cite{wkyp,mvm} analyzed the canonical structure of the NCCB in the framework of the 
	BFT embedding method, aiming to promote the NCCB constraints to first-class. The BFT formalism \cite{A_10,A_11,A_12} converts second-class constraints into first-class ones by enlarging the phase space with the introduction of auxiliary fields and has found many important applications in the literature from which we mention a short representative sample \cite{Amorim:1995sh, bft1,bft2, Amorim:1999xr, bft3, DeAbreu:2000oil, bft4, Mandal:2022xuw}.
	In \cite{wkyp}, it is shown that the direct application of the BFT embedding formalism to the NCCB model may lead to an infinite amount of auxiliary fields in phase space.  With an alternative choice for the BFT fields symplectic structure, Majid, Vahid and Mehran have shown in \cite{mvm} that it is possible to abelianize the NCCB model in finite order, reducing the number of auxiliary fields to only two.  Nonetheless, we claim that a consistent NCCB abelianization can be done without the need of any auxiliary fields whatsoever.  That is one of the main advantages of the gauge-unfixing (GU) method \cite{mr, vyt1,vyt2,vyt3, A_14, db, epl1, epl2, proto}.  Building on the original work of Mitra and Rajaraman \cite{mr}, which first conjectured the interpretation of second-class constraints in phase space as resulting from gauge-fixing conditions within a larger gauge invariant theory, Anishetty and Vytheeswaran constructed a Lie projection operator whose action in the second-class functions was able to reveal hidden symmetries \cite{vyt1,vyt2,vyt3}.  Those ideas were further elaborated to produce the improved or modified gauge-unfixing formalism \cite{A_14, db}, centered on the construction of the invariant GU variables, and have found recent appeal in quantum field theory \cite{epl1, epl2, proto}. In this way, the main goal of the present letter is to develop a gauge invariant theory for the NCCB model, without auxiliary fields, by directly applying the modified GU formalism \cite{A_14, db, epl1,epl2, proto} to the noncommutative fields space. 
	
	For the reader's convenience, we have organized our presentation as follows: In the next section, we discuss the NCCB model and analyse its constraints canonical structure with the use of the Dirac-Bergmann formalism \cite{Dirac:1950pj, Anderson:1951ta, Dirac, Sundermeyer:1982gv} computing the Dirac brackets algebra in phase space. In Section {\bf 3}, we take the opportunity to briefly review the modified GU formalism preparing its application to the NCCB and turning the article self-contained.  In Section {\bf 4}, the modified GU formalism is applied to the NCCB model and we show that it is possible to obtain gauge invariance without the introduction of auxiliary fields. We close in the last section with our conclusions and final remarks. 
	
	\section{The Noncommutative Chiral Bosons Model}
	
	The noncommutative chiral bosons model (NCCB) in $(1+1)$ space-time dimensions is defined by the first-order action \cite{dgms}
	\begin{equation}
		S[\phi_a] = \int d^2x \left[ -\frac{2}{1+\theta^2} \dot{\phi}_a\Delta_{ab}\phi_b' - \phi_a'\phi_a'\right] \,,
		\label{S}
	\end{equation}
	where $\theta$ denotes a noncommutative parameter and $\Delta_{ab}$ is an invertible symmetric $2\times 2$ matrix given by
	
	\begin{equation*}
		\Delta_{ab} = \frac{1}{2}
		\begin{pmatrix}
			-1 & \theta \\
			\theta & 1  
		\end{pmatrix}
		\,.
	\end{equation*}
	The Latin indexes $a,b$ run through $1,2$ and   
	the dynamics resulting from (\ref{S}) describes two chiral bosons $\phi_1$ and $\phi_2$ coupled by the noncommutative 
	parameter $\theta$.  In fact, the field equations directly derived from the minimum action principle applied to $S[\phi_a]$ read
	\begin{equation}
		2\frac{\Delta_{ab}}{1+\theta^2}\dot{\phi}_b'+\phi_a'' = 0 \,,
	\end{equation}
	and, alternatively, can be cast into the form
	\begin{equation}
		\dot{\phi}_a' + 2 \Delta_{ab}\phi_b'' = 0 
		\,.
		\label{EOM2}
	\end{equation}
	In the limit $\theta\rightarrow 0$, the left and right modes decouple and we recover the usual commutative case consisting of two independent chiral bosons
	\cite{fj, Amorim:1995tk}.  It is interesting to notice that the non-commutative parameter $\theta$ leads to an enhancement of the speed of light concerning the two chiral bosons propagation.  This can be seen from the fact that the equations of motion (\ref{EOM2}) are equivalent to the pair
	\begin{equation}
		\begin{aligned}
			\dot{\phi}_1' &= ~\phi_1''-\theta \phi_2''\,,\\
			\dot{\phi}_2' &= -\phi_2''-\theta \phi_1''\,,
		\end{aligned}
	\end{equation}
	which in turn, by space integration and time derivation, result in
	\begin{equation}
		\Box_\theta \phi_1 = \Box_\theta \phi_2 = 0
		\,,
	\end{equation}
	where $\Box_\theta$ denotes the $\theta$-dependent noncommutative  D'Alembertian operator defined as
	\begin{equation}
		\Box_\theta\equiv \frac{1}{1+\theta^2}\partial_t^2 - \partial_x^2
		\,.
	\end{equation}
	Thus, we see that Lorentz invariance is preserved in noncommutative space, as long as we redefine the speed of light $c$ as
	\begin{equation}
		c\rightarrow c_\theta \equiv c\sqrt{1+\theta^2}\,.
	\end{equation}
	
	Due to the presence of constraints, the canonical quantization of the NCCB must be done carefuly. The action (\ref{S}) actually corresponds to a singular Dirac-Bergmann system \cite{Dirac:1950pj, Anderson:1951ta, Dirac, Sundermeyer:1982gv}. To see this feature, note that the canonical momenta in phase space can be written as
	\begin{equation}
		\label{cm}
		\pi_a=-\frac{2}{1+\theta^2}\Delta_{ab}\phi_b' \,.
	\end{equation}
	As we can see, Eq. (\ref{cm}) does not involve the fields time derivatives. Consequently, we have a pair of primary constraints in phase space given by
	\begin{equation}
		\label{pc}
		\Omega_a \equiv \pi_a + \frac{2}{1+\theta^2}\Delta_{ab}\phi_b' \,,
	\end{equation}
	with corresponding Poisson bracket relations
	\begin{equation}
		\label{pb}
		\left\{
		\,\Omega_a(x)\,,\,\Omega_b(y)\,
		\right\}= \frac{4}{1+\theta^2}\Delta_{ab}\delta'(x-y) \,.
	\end{equation}
	
	Concerning the dynamical evolution,
	the Legendre transformation from configuration to phase space leads to a well defined canonical Hamiltonian within the primary constraints hypersurface given by
	\begin{equation}
		\label{sech}
		H = \int dx \,\phi_a' \phi_a'
	\end{equation}
	and further steps of the Dirac-Bergmann algorithm show the consistent stability of $\Omega_a$  without the need of new constraints. Hence, the constraints set (\ref{pc}) is complete and the invertibility of Eq. (\ref{pb}) assures the second-class nature of the system.  To obtain the Dirac brackets among the phase space variables, we note that the antisymmetric inverse of (\ref{pb}) can be written as
	\begin{equation}
		\label{invpb}
		\left\{
		\,\Omega_a(x)\,,\,\Omega_b(y)\,
		\right\}^{-1}= \Delta_{ab}\epsilon (x-y) \,,
	\end{equation}
	with $\epsilon(x)$ denoting the antisymmetric unity step function 
	satisfying
	\begin{equation}
		\epsilon (x) = -\epsilon (-x) \,,
	\end{equation}
	and
	\begin{equation}
		\epsilon'(x) = \delta(x) \,.
	\end{equation}
	Eq. (\ref{invpb}) represents the functional inverse of (\ref{pb}) in the sense of
	\begin{equation}
		\int dz \,
		\left\{
		\,\Omega_a(x)\,,\,\Omega_c(z)\,
		\right\}
		\left\{
		\,\Omega_c(z)\,,\,\Omega_b(y)\,
		\right\}^{-1} = \delta_{ab}\delta(x-y) \,,
	\end{equation}
	and
	\begin{equation}
		\int dz \,
		\left\{
		\,\Omega_a(x)\,,\,\Omega_c(z)\,
		\right\}^{-1}
		\left\{
		\,\Omega_c(z)\,,\,\Omega_b(y)\,
		\right\}
		= \delta_{ab}\delta(x-y)
		\,.
	\end{equation}
	Inserting (\ref{invpb}) into the general DB definition
	\begin{equation}
		\{F,G\}_D = \{F,G\}-\int dz d\bar{z} \, \{F,\Omega_a(z)\}\{\Omega_a(z),\Omega_b(\bar{z})\}^{-1}\{\Omega_b(\bar{z}),G\}
		\,,
	\end{equation}
	the fundamental DBs among the phase space variables can be readily computed as
	\begin{equation}
		\{\phi_a(x),\phi_b(y)\}_D = \Delta_{ab} \epsilon(x-y)\,,
	\end{equation}
	\begin{equation}
		\{\phi_a(x),\pi_b(y)\}_D = \frac{1}{2}\delta_{ab} \delta(x-y) \,,
	\end{equation}
	and
	\begin{equation}
		\{\pi_a(x),\pi_b(y)\}_D = -\frac{1}{1+\theta^2}\Delta_{ab} \partial_x \delta(x-y)\,.
	\end{equation}
	At this point,
	the canonical quantization can be pursued by requiring the associated operators to satisfy commutation relations dictated by the DB algebra above.  Our main goal in the present work, however, is to produce gauge symmetry for the NCCB model, write down the corresponding quantum generating functional, and proceed along the lines of the functional quantization framework.  This can be done by means of the modified GU formalism. 
	Indeed, the second-class property shown by the
	constraints in Eq. (9) allows us to directly apply that improved version of the GU formalism considering one of the constraints as generator of gauge transformations and calculating the GU variables.  In the next section, we give a brief general review of the modified GU technique, paving the way for its application to the NCCB in the following one.
	
	\section{Brief Review of the modified GU formalism}
	\label{sqm}
	The modified gauge unfixing formalism developed by Neto \cite{A_14,db,mes,proto} is based on the idea of selecting one of the two second-class constraints to be the gauge symmetry generator and the other one being discarded in a broader gauge-invariant context. 
	Consider for example a given second-class phase-space function $T(A_{\mu},\pi_{\mu})$ with the index $\mu$ running through all phase space variables. Our strategy is to write a first-class function $\tilde{T}({A}_{\mu},{\pi}_{\mu})$ obtained from the second-class function $T$ as
	\begin{equation}
		\tilde{T}({A}_{\mu},{\pi}_{\mu}) \equiv {T}(\tilde{A}_{\mu},\tilde{\pi}_{\mu})
		\,,
	\end{equation}
	by redefining the original phase space variables
	\begin{equation}
		A_\mu \longrightarrow \tilde{A}_\mu ({A}_{\mu},{\pi}_{\mu}) \,,
	\end{equation}
	\begin{equation}
		\pi_\mu \longrightarrow \tilde{\pi}_\mu ({A}_{\mu},{\pi}_{\mu}) \,,
	\end{equation}
	such that
	\begin{align}
		\delta \tilde{A}_\mu = \alpha \left \{ \tilde{A}_\mu , \psi \right \}=0 \,, \label{3.8}
	\end{align}
	and
	\begin{align}
		\delta \tilde{\pi}_\mu = \alpha \left \{ \tilde{\pi}_\mu , \psi \right \}=0 \,, 
	\end{align}
	where $\alpha$ is an infinitesimal parameter and $\psi$ is the second-class constraint chosen to be the gauge symmetry generator. The deformed variables $\tilde{A}_\mu$, $\tilde{\pi}_\mu$ are known as {\it GU variables}. It is clear now that functions of the GU variables, in particular $\tilde{T}$, will be gauge invariant since
	\begin{align}
		\left \{ \tilde{T}, \psi \right \}= \left \{ \tilde{A}, \psi \right \} \frac{\partial  {T} }{\partial \tilde{A}} + \frac{\partial {T}}{\partial \tilde{\pi }}\left \{ \tilde{\pi }, \psi \right \} = 0 \,.
	\end{align}
	Consequently, we can obtain a gauge invariant function from the replacement of
	\begin{align}
		\label{repla}
		T(A_{\mu}, \pi_{\mu}) \rightarrow T(\tilde{A_{\mu}}, \tilde{ \pi_{\mu}})=\tilde{T}({A_{\mu}},{ \pi_{\mu}}) \,.
	\end{align}
	
	Now suppose the system has only two second class constraints $Q_1$ and $Q_2$. So, the GU gauge invariant phase space variables, collectively denoted by $\tilde{\Lambda}\equiv (\tilde{A}_\mu,\tilde{\pi}_\mu)$, can be constructed as a power the series in the discarded constraint $Q_{2}$
	\begin{align}
		\tilde{\Lambda}(x)=\Lambda(x)+\int dyb_{1}(x,y)Q_{2}(y)+\iint dydz b_{2}(x,y,z)Q_{2}(y)Q_{2}(z)+... \,, \label{3.11}
	\end{align}
	satisfying, on the constraint surface $Q_{2}=0$, the boundary condition 
	\begin{align}
		{\tilde{\Lambda}}_{\big| Q_{2}=0}=\Lambda \,.
	\end{align}
	This assures that we recover the original second-class system when $Q_2=0$. The coefficients $b_{n}$ in relation~\eqref{3.11} are then determined by the GU invariant requirement
	\begin{equation}
		\delta\tilde{\Lambda}= \alpha \left \{ \tilde{\Lambda} , Q_1 \right \}=0\,.
	\end{equation}
	The general equation for $b_{n}$ is
	\begin{align}
		\delta \tilde{\Lambda}(x)=\delta \Lambda(x)+\delta \int dyb_{1}(x,y)Q_{2}(y)+ \delta \iint dydz b_{2}(x,y,z)Q_{2}(y)Q_{2}(z)+...=0 \,, \label{3.18}
	\end{align}
	in which we have
	\begin{align}
		& \delta \Lambda(x)=\int dy \, \alpha(y)\left \{ \Lambda(x), \psi(y) \right \}, \\
		&\delta b_{1}(x)=\int dy \, \alpha(y)\left \{ b_{1}(x), \psi(y) \right \},\\
		&\delta Q_{2}(x)=\int dy \, \alpha(y)\left \{ Q_{2}(x), \psi(y) \right \} \,.
	\end{align}
	So, for the first order correction term $(n = 1)$ we have from Eq.~\eqref{3.18} 
	\begin{align}
		\label{b1s}
		\delta \Lambda(x)+\int dyb_{1}(x,y)\delta Q_{2}(y)= 0 \,.
	\end{align}
	From Eq. ({\ref{b1s})} we can determine the coefficient $b_1$.
	For the second order correction term $(n = 2)$ we have
	\begin{eqnarray}
		\label{b2s}
		\int dy\delta b_{1}(x,y)Q_{2}(y)+2\iint dydz b_{2} (x,y,z)\delta Q_{2}(y)Q_{2}(z)= 0 \,.
	\end{eqnarray}
	Then, from Eq. ({\ref{b2s})} we can determine the coefficient $b_2$ and so on and so forth.
	Therefore, from the GU variables power series defined in Eq.~(\ref{3.11}) we can derive a corresponding gauge invariant theory.
	
	\section{Disclosing hidden symmetries}
	In this section, we apply the modified GU formalism to the NCCB. As we have seen, the NCCB has two second-class constraints given by Eq. (\ref{pc}). Writting them out explicitly in terms of the components of $\Delta_{ab}$, we have
	\begin{eqnarray}
		\label{omega1}
		\Omega_1 = \pi_1 - \frac{\phi'_1}{1+\theta^2} + \frac{\theta\, \phi'_2}{1+\theta^2} \,, \\
		\label{omega2}
		\Omega_2 = \pi_2 + \frac{\theta \,\phi'_1}{1+\theta^2} + \frac{\phi'_2}{1+\theta^2}    \,.
	\end{eqnarray}
	Then, following the ideas of the modified GU formalism, one of the two second-class constraints will be chosen to be the gauge symmetry generator
	and the other one will be discarded. Thus, two possible choices for the gauge symmetry generator for the NCCB are possible. We consider below each of these
	two different cases separately.
	\subsection{Case A}
	In this first case, we consider the constraint $\Omega_1$, Eq. (\ref{omega1}), as the gauge symmetry generator and discard the other second-class constraint $\Omega_2$, Eq. (\ref{omega2}). The $\varepsilon$-infinitesimal gauge transformations generated by $\Omega_1$ 
	are given by
	\begin{align}
		\label{dphi1}
		\delta \phi_1 &= \int dy\, \alpha(y) \, \{\phi_1 , \Omega_1(y)\} = \alpha \,,\\
		\label{dfieldphi2}
		\delta \phi_2 & = \int dy\, \alpha(y) \, \{\phi_2 , \Omega_1(y)\} = 0 \,,
		\\
		\label {dpi1}
		\delta \pi_1 &= \int dy\, \alpha(y) \, \{\pi_1 , \Omega_1(y)\} = -\frac{1}{1+\theta^2} \, \alpha' \,,\\
		\label {dpi2}
		\delta \pi_2 &= \int dy\, \alpha(y) \, \{\pi_2 , \Omega_1(y)\} = \frac{\theta}{1+\theta^2} \, \alpha' \,,
	\end{align}
	while the constraint $\Omega_2$ transforms under $\Omega_1$ as
	\begin{align}
		\label{domega2}
		\delta \Omega_2 &= \int dy \, \alpha(y) \, \{\Omega_2,\Omega_1 (y) \} =\frac{2 \theta}{1+\theta^2} \,  \alpha' \,.
	\end{align}
	Since the fields $\phi_1, \pi_1$ and $\pi_2$ are not naturally gauge invariant under $\Omega_1$, following the improved GU approach, we need to construct their GU invariant combinations in terms of a power series in $ \Omega_2 $.  For instance, the first gauge invariant GU variable $\tilde{\phi}_1$ can be written as
	\begin{eqnarray}
		\label{fi1}
		\tilde{\phi}_1(x) = \phi_1(x) + \int dy b_1(x,y) \,\Omega_2(y) + \int dy dz b_2(x,y,z) \,\Omega_2(y) \Omega_2(z) + ... \,,
	\end{eqnarray}
	with the correction coefficient functions $b_n$ to be calculated from the invariant condition $ \delta \tilde{\phi}_1 = 0$. For the linear correction
	term $(n=1)$, we have
	\begin{eqnarray}
		\label{dfi11}
		\delta \phi_1 + \int dy  b_1(x,y) \delta \Omega_2(y) = 0  \,,
	\end{eqnarray}
	and,
	plugging Eqs. (\ref{dphi1}) and (\ref{domega2}) into (\ref{dfi11}), we obtain
	\begin{eqnarray}
		\label{b1}
		b_1(x,y) = - \frac{1+\theta^2}{2\theta}\epsilon(x-y) \,.
	\end{eqnarray}
	For the quadratic term we have $b_2 = 0$ and thus, for $ n \geq 2$, all the remaining correction coefficient functions $b_n$ are null.
	Taking this into consideration and inserting Eq. (\ref{b1}) in (\ref{fi1}), we obtain the final expression for the first gauge-invariant 
	GU variable $\tilde{\phi}_1$ as
	\begin{equation}
		\label{phi1g}
		\tilde{\phi}_1(x) = \phi_1(x) - \frac{1+\theta^2}{2\theta} \int dy \epsilon(x-y)\Omega_2(y) \,.
	\end{equation}
	Repeating this same iterative process, in a similar fashion, we can derive the remaining gauge-invariant GU variables in phase space as
	\begin{eqnarray}
		\label{phi2g}
		\tilde{\phi}_2 &=& \phi_2 \,, \\
		\label{pi1g}
		\tilde{\pi}_1 &=& \pi_1 + \frac{1}{2\theta} \, \Omega_2 \,, \\
		\label{pi2g}
		\tilde{\pi}_2 &=& \pi_2 - \frac{1}{2} \, \Omega_2 \,.
	\end{eqnarray}
	Differentiating with respect to space, from Eqs. (\ref{phi1g}) and (\ref{phi2g}), we have the further useful relations
	\begin{eqnarray}
		\tilde{\phi}'_1 &=&\phi'_1 - \frac{1+\theta^2}{2\theta} \Omega_2\,,\\
		\tilde{\phi}'_2 &=&\phi'_2 \,.
	\end{eqnarray}
	
	By construction, as seen in the previous section, any function of the GU variables is automatically gauge-invariant.
	In particular, the gauge-invariant GU Hamiltonian can be directly obtained from Eq. (\ref{sech}) as
	\begin{eqnarray}
		\label{guh}
		\tilde{H} =  \int dx \left[
		\left(\phi'_1 - \frac{1+\theta^2}{2\theta} \, \Omega_2 \right)^2 + \phi'_2 \phi'_2 
		\right]
		\,,
	\end{eqnarray}
	and satisfies the gauge-invariant condition $ \{ \tilde{H}, \Omega_1 \}  = 0$.  Due to its gauge invariance, the GU Hamiltonian can be used to 
	obtain the phase space partition function in the functional quantization approach \cite{fad} as 
	\begin{equation}\label{Z}
		Z = \int {\cal{D}} \phi_1 {\cal{D}} \pi_1 {\cal{D}} \phi_2 {\cal{D}} \pi_2  \, \delta(\Omega_1) \delta(\Gamma_1)
		\, det |\{ \Omega_1, \Gamma_1\}| e^{i S} \,,
	\end{equation}
	where $\Gamma_1$ denotes a suitable gauge-fixing condition and
	\begin{eqnarray}
		S = \int d^2x   \left( \pi_1 \dot{\phi}_1 + \pi_2 \dot{\phi}_2 - \tilde{H} \right) \,.
	\end{eqnarray}
	Concerning gauge achievability,
	the gauge fixing function $\Gamma_1$ must be chosen in such a way that
	the determinant in the integration measure in (\ref{Z}) does not vanish.  This concludes the functional quantization of the gauge-invariant description of the NCCB, with gauge transformations generated by $\Omega_1$.
	
	\subsection{Case B}
	Next, we consider the other possible choice for the gauge symmetry generator. Let now $\Omega_2$ generate the gauge symmetries, certainly different from the ones in case A, while the constraint $\Omega_1$ is discarded. The infinitesimal gauge transformations generated by $\Omega_2$ read
	\begin{align}
		\label{dfi12}
		\delta \phi_1 &= 0 \,,\\
		\label{dfi22}
		\delta \phi_2 &= \int dy \, \alpha(y) \, \{\phi_2 , \Omega_2(y)\} = \alpha \,,\\
		\label {dpi1b}
		\delta \pi_1 &= \int dy \, \alpha(y) \, \{\pi_1 , \Omega_2(y)\} = \frac{\theta}{1+\theta^2} \alpha' \,,\\
		\label {dpi2b}
		\delta \pi_2 &= \int dy \, \alpha(y) \, \{\pi_2 , \Omega_2(y)\} =\frac{1}{1+\theta^2} \alpha' \,,\\
		\label{domega1}
		\delta \Omega_1 &= \int dy \, \alpha(y) \, \{\Omega_1,\Omega_2(y)\} = \frac{2\theta}{1+\theta^2} \, \alpha' \,.
	\end{align}
	Eq. \eqref{dfi12} shows that $\phi_1$ is already gauge invariant under transformations generated by $\Omega_2$, thus
	\begin{equation}
		\tilde{\phi}_1=\phi_1 \,. \nonumber
	\end{equation}
	The gauge invariant field $\tilde{\phi_2}$ is constructed as
	\begin{eqnarray}
		\label{fi2}
		\tilde{\phi}_2 = \phi_2 + \int dy c_1(x,y) \,\Omega_1(y) + \int dydz c_2(x,y,z) \,\Omega_1(y) \Omega_1(z) + ... \,.
	\end{eqnarray}
	Now, imposing the gauge-invariant condition $ \delta \tilde{\phi}_2 = 0$, the correction terms $c_n$ can be obtained. For the linear correction term $(n=1)$, we have
	\begin{eqnarray}
		\label{dfi21}
		\delta \phi_2(x) + \int dy c_1(x,y) \delta \Omega_1(y) = 0  \,.
	\end{eqnarray}
	Using Eqs. \eqref{dfi22} and \eqref{domega1} we find
	\begin{eqnarray}
		\label{c1}
		c_1(x,y) = - \frac{1+\theta^2}{2 \theta} \epsilon(x-y) \,.
	\end{eqnarray}
	It is easy to see that the variation of $c_1$ leads to $c_2 = 0$. Then, for $ n \geq 2$, all correction terms $c_n$ are null. 
	Therefore, by putting Eq. \eqref{c1} into Eq. \eqref{fi2}, the GU variable $\tilde{\phi}_2$ acquires the form 
	\begin{eqnarray}
		\tilde{\phi}_2 = \phi_2 - \frac{1+\theta^2}{2\theta} \int dy \epsilon(x-y)\Omega_1(y) \,. \nonumber
	\end{eqnarray}
	The remaining GU variables can be obtained by the same iterative process. Proceeding this way and putting all GU variables together we have
	
	\begin{align}
		\label{fi1g1}
		\tilde{\phi}_1 &= \phi_1  \,,\\
		\label{fi2g2}
		\tilde{\phi}_2 &= \phi_2 - \frac{1+\theta^2}{2\theta} \int dy \epsilon(x-y)\Omega_1(y) \,,\\
		\label{tildepi1b}
		\tilde{\pi}_1 &= \pi_1 - \frac{1}{2} \, \Omega_1 \,,\\
		\label{tildepi2b}
		\tilde{\pi}_2 &= \pi_2 - \frac{1}{2\theta} \, \Omega_1 \,.
	\end{align}
	From Eqs. (\ref{fi1g1}) and (\ref{fi2g2}), it immediately follows 
	\begin{eqnarray}
		\label{fil12}
		\tilde{\phi}'_1 &=& \phi'_1 \,\\
		\label{fil22}
		\tilde{\phi}'_2 & =& \phi'_2 - \frac{1+\theta^2}{2\theta} \, \Omega_1 \,.
	\end{eqnarray}
	Now we are able to write the gauge invariant version of the Hamiltonian \eqref{sech}. Substituting Eqs. \eqref{fil12} and \eqref{fil22} into \eqref{sech} we have
	\begin{eqnarray}
		\label{guh2}
		\tilde{H} =  \int dx \left[
		\left(\phi'_2 - \frac{1+\theta^2}{2\theta} \, \Omega_1 \right)^2 + \phi'_1 \phi'_1 
		\right]
		\,.
	\end{eqnarray}
	By construction, the GU Hamiltonian, $\tilde{H}$, satisfies the condition $ \{ \tilde{H}, \Omega_2 \}  = 0$.
	
	As was done in case A, given the gauge invariant Hamiltonian, we can derive the phase space partition function
	\begin{eqnarray}
		Z = \int {\cal{D}} \phi_1 {\cal{D}} \pi_1 {\cal{D}} \phi_2 {\cal{D}} \pi_2  \, \delta(\Omega_2) \delta(\Gamma_2)
		\, det |\{ \Omega_2, \Gamma_2\}| e^{i S} \,,
	\end{eqnarray}
	where
	\begin{eqnarray}
		S = \int d^2x   \left( \pi_1 \dot{\phi}_1 + \pi_2 \dot{\phi}_2 - \tilde{H} \right) \,,
	\end{eqnarray}
	with the Hamiltonian density $\tilde{H}$ given by Eq. (\ref{guh2}). The gauge fixing condition $\Gamma_2$ is chosen so that
	the determinant appearing in the functional measure is nonvanishing.  In both cases A and B, we were able to obtain a gauge-invariant version for the NCCB model.
	
	\section{Conclusions}
	In this work, we have converted the NCCB model into a first-class constrained system using a modified GU formalism, whose convenience lies on the freedom of choice of the gauge symmetry generator and on the redefinition of the phase space itself without using any extra variables. One of the constraints becomes the generator of gauge symmetries and the other one is discarded. Such ambiguity allowed us to obtain two gauge invariant systems consistent with the original second-class one, which can be recovered back in a straightforward manner by setting the discarded constraint equal to zero. In case A, the constraint $\Omega_1$, Eq. \eqref{omega1}, was chosen in order to be the gauge symmetries generator, while in case B the constraint $\Omega_2$, Eq. \eqref{omega2}, was selected to be the generator of gauge symmetries. We can note that the canonical structure acquired from the modified GU formalism, for both cases, is similar to those obtained from other approaches. It is worth mentioning the non-local form derived for the fields $\tilde{\phi}_1$, in case A, and $\tilde{\phi}_2$, in case B, characterized by the presence of the antisymmetric step function $\epsilon(x-y)$ in their expressions, since non-locality can also be generated from noncommutative field theories. Here, we can state that the GU variables are dictating the rules for obtaining a gauge theory from a second-class constrained system. Thus, as has become clear throughout our work, once GU variables are computed the corresponding gauge theory is obtained consistently and in a very simple way.
	\label{FR}
	
	\section{Acknowledgments}
	The authors sincerely thank Nikoofard Vahid for useful comments. RT kindly thanks the CERN Theoretical Physics Department (CERN-TH) for hospitality in a insightful research environment.  The CAPES (Coordenação de Aperfeiçoamento de Pessoal de Nível Superior) and FAPEMIG (Fundação de Amparo à Pesquisa do Estado de Minas Gerais) are ackknowledged for financial support. Jorge Ananias Neto thanks CNPq (Conselho Nacional de Desenvolvimento Cient\'ifico e Tecnol\'ogico), Brazilian scientific support federal agency, for partial financial support, CNPq-PQ, Grant number 307153/2020-7.

\end{document}